\documentclass[a4]{article}
\usepackage[dvips]{graphicx}

\begin{document}

\title{Towards a model of incoherent scatter signal spectra without averaging 
over sounding runs} 

\author{O. I. Berngardt}
\date{}
\maketitle
{\par\centering
Institute of Solar-Terrestrial Physics,

Irkutsk, Russia, 664033.

Lermontova Str.,126, PBox 4026,

(berng@iszf.irk.ru)
\par}

\begin{abstract}
This paper offers a model for incoherent scatter signal spectra without 
averaging the received signal over sounding runs (realizations). The model 
is based on the existent theory of radio waves single scattering from the 
medium dielectric permittivity irregularities, and on the existing kinetic 
theory of the plasma thermal irregularities. The proposed model is obtained 
for the case of monostatic sounding. The model shows that the main 
contribution to the received signal is made by ion-acoustic waves caused by 
certain spatial harmonics of the ions number phase density. The model 
explains the width and form of the signal spectrum by macroscopic 
characteristics of the medium, and its fine 'peaked' structure by 
characteristics of the ions number phase density. The notion of the weight
volume is introduced to define the domain of wave vectors and velocities in
the spatial spectrum of the ions number phase density which makes the main 
contribution to the formation of the scattered signal. This weight volume 
depends on the antenna pattern, the form of the sounding signal, and on the 
time window of spectral processing, as well as on ionospheric plasma 
macroscopic parameters: electron and ion temperatures, ion composition, 
and the drift velocity. Within the context of additional assumption about 
the ions number phase density, the proposed model was tested by the data 
from the Irkutsk incoherent scatter radar. The test showed a good fit of 
the model to experiment. 
\end{abstract}


\section{Introduction}

One of the remote probing techniques for the ionosphere is the method of 
radio waves incoherent scatter. 
The method is based on the scattering of radio waves from ionospheric plasma 
dielectric permittivity irregularities [{\it Evans}, 1969].
Furthermore, two different experimental configurations are involved: monostatic
(where the receive and transmit antennas are combined) and bistatic (where 
these antennas are spaced). In actual practice, it is customary to use the 
monostatic configuration. Ionospheric plasma parameters (ion composition, 
drift velocity,
electron and ion temperatures, and electron density) in this case are determined
from the scattered signal received after completion of the radiated pulse. The
spectral power of the received signal, averaged over sounding runs ('realizations'),
is related (assuming that such an averaging is equivalent to statistical averaging)
to the mean spectral density of dielectric permittivity irregularities by the
radar equation [{\it Tatarsky}, 1969]. The connection of the dielectric 
permittivity irregularities spectral density  with mean statistical parameters 
of the medium is usually determined in terms of kinetic theory 
[{\it Clemow and Dougherty},1969; {\it Sheffield}, 1975; {\it Kofman},1997]. 

The location and size of the ionospheric region that makes a contribution to
the scattered signal (sounding volume) is determined by the antenna beam shape,
the sounding radio pulse, and by the time window of spectral processing 
[{\it Suni et al.}, 1989].
The shape of the sounding volume determines also the method's spectral resolution,
the accuracy to which the mean spectral density of dielectric permittivity is
determined (which, in turn, affects the determination accuracy of macroscopic
ionospheric parameters: electron and ion temperatures, the drift velocity, and
electron density). The number of realizations, over which the received signal
spectral power is averaged, determines the method's time resolution, i.e. its
ability to keep track (based on measurements) of fast changes of macroscopic
parameters of ionospheric plasma.

Currently most incoherent scatter radars have accumulated extensive sets of
the scattered signal individual realizations  (private communications of 
P.Erickson (Millstone Hill), V.Lysenko (Kharkov IS radar), and G.Wannberg 
(EISCAT)). Therefore, attempts are made to analyze the realizations from 
different methods which differ from a standard averaging by their sounding 
runs. Basically, these methods imply looking for small scatterers making the 
main contribution to the scattered signal. This method is good for analyzing 
signals scattered from meteors and their traces [{\it Pellinen-Wannberg}, 1998]; 
however, it is insufficiently substantiated for describing the scattering in 
the ionosphere. 

In the work there were used the experimental data obtained with Irkutsk 
Incoherent Scatter radar. The radar is located at \( 52^{0}N,104^{0}E \), it 
has sounding frequency 152-160 MHz and peak power 3MW. High signal-to-noice ratio 
during  the experiments under investigation ( S/N > 10 ) allows us to neglect the 
noice effects when analyzing the signal received. 

The technique of the incoherent scatter signal processing in Irkutsk IS radar 
is the following. For each single realization of received signal we 
calculate spectrum in time window with width equal to the sounding signal 
duration and with delay corresponding to the radar range to the sounding 
volume investigated. 
The sounding signal we use in this experiment is a radiopulse with 
duration 800 mks. The repeating frequency approximately 25 Hz. 
Averaging over the 1000 realizations corresponds to \( 3 \% \) 
dispersium of the overaged spectrum relative to its mathematical expectation.  
The reason of using such a simple pulse is to investigate the fine structure 
of the single (unaveraged) spectrums in this simpliest case.

\begin{figure*}
\vskip1in  

\resizebox*{0.9\textwidth}{!}{
\includegraphics{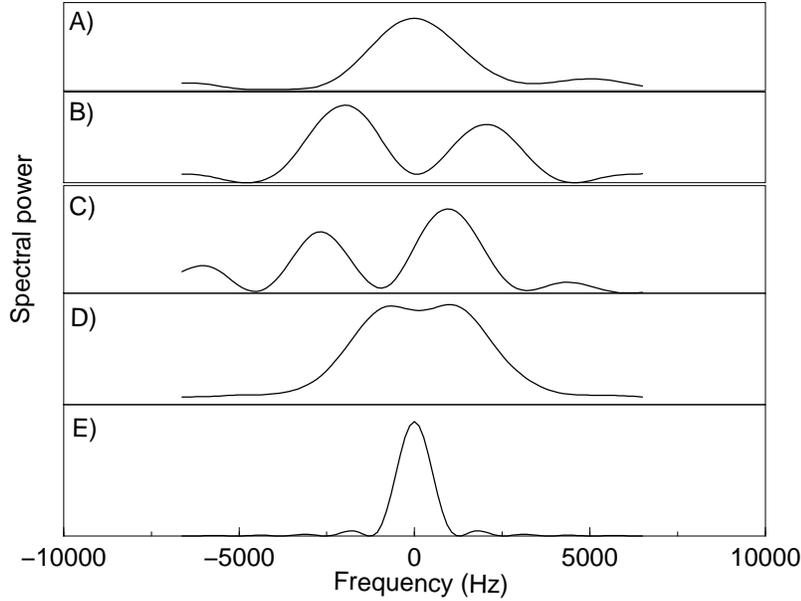} 
}
\caption{The spectral power of three different realizations (A-C), \
averaged over 1000 realizations spectral power of the scattered \
signal (D) and spectral power of the sounder signal envelope (E) as \
deduced using the data from the Irkutsk incoherent scatter radar.} 

\label{figone}
\end{figure*}

Figure~\ref{figone} exemplifies the mean spectral power of the 
scattered signal and its separate realizations, based on the data from the 
Irkutsk Incoherent Scatter radar. It
is evident from the figure that the spectral power of the scattered signal in
an individual realization (Figure~\ref{figone}(b-d)) differs drastically 
from that averaged over realizations (Figure~\ref{figone}(a)); 
therefore, existing model of the incoherently scattered signal, based on averaging 
over sounding runs, are inapplicable for
its interpretation. For that reason, development of new models of the scattered
signal for analyzing its separate realizations without averaging them is important
from the theoretical and practical standpoint. 

Sometimes it is useful to suppose that incoherent scattering signal is a random 
gaussian one [for example, {\it Farley}, 1969; {\it Zhou}, 1999]. 
But, it is well known that the signal received is a detirministic 
function of ionospheric dielectric permittivity \( \epsilon \) and is fully 
determined in first approximation by the Born's formula (in one or another 
its form [{\it Ishimaru}, 1978; {\it Berngardt and Potekhin}, 2000]), 
this relation could be called as a radar equation for signals
[{\it Berngardt and Potekhin}, 2000]. 

The dielectric permittivity irregularities also could be supposed as a random 
functions, but they are deterministic functional of some other functions 
(in case of uncollisional unmagnetized palsma with one ions type those 
functions are phase density of the ions and electrons as functions of velocity, 
location and time, ion composition and temperatures of the ions and electrons,
this functional dependence is determined by the Landau's solution 
[{\it Landau}, 1946]). 

If one could determine all these unknown functions, the received 
signal shape in single realization will be fully determined, and could be 
analyzed without using any statistical methods. Such an approach, for example,
is used in radioacoustical technique of the athmosphere sounding when 
the delectric permittivity irregularities (by which the radisignal is scattered)
are generated by the acoustical wave [{\it Kalistratova and Kon}, 1985].

\begin{figure*}
\vskip1in  
\resizebox*{0.9\textwidth}{!}{
\includegraphics{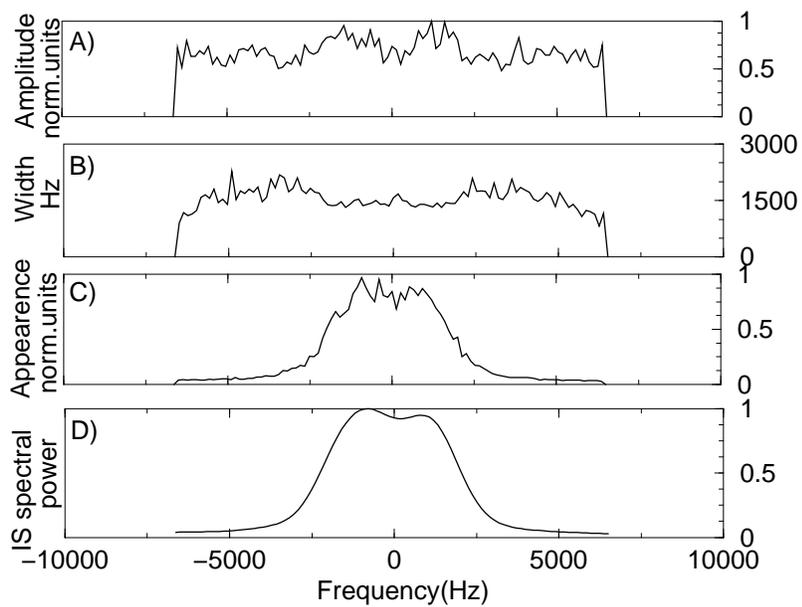} 
}
\caption {Statistical proporties of single spectrum realizations (averaged over 
 1000 realizations) as functions of frequency. Mean peak amplitude (A),
 mean peak width(B) and mean appearence of the peaks (C) for different 
 frequencies of received signal. For comparison there is showed the IS 
 spectral power for this data (D).}

\label{figstat}
\end{figure*}

The statistical proporties of the single realizations are showed 
at Figure~\ref{figstat}. From this figure it becomes 
clear that the unaveraged spectrum has the fine structure - it consists from
a number of peaks with approximately 1.5KHz width (and this width very slightly 
depends on frequency), which could be characterized by the peak 
amplitude(amplitude at the maximum of the peak) and peak appearence (number of realizations 
in which there is a peak maximum at given frequency) at the given frequency, 
and those properties distributions are not gaussian ones but have double peaked 
structure and located in the same band with incoherent scattering average spectral 
power. 

This fact allows as to suppose that not only average spectral power of the 
received signal depends on ionosperical parameters, but the fine structure of 
non-averaged spectra too. 

At first, it is neccessary to understand quilitatively, what information 
one could obtain from one realization of the IS signal. It is well known, 
that after any statistical processing of a function a part of the information 
is loosed irreversibly (for example. when one calucaltes the first n statistical 
moments, all the rest moments, starting with n+1 are still unknown). That is why, 
if the statistical characteristics of the realizations (mean spectral power or 
correlation function) are depend on the ion and electron temperatures and the ion 
composition then single realization must depend on all those parameters and 
on some new 'additional' parameters. It is clear that to determine 
temperatures and ion composition from averaged signal parameters is much easier 
than from single realization (because the second one includes additional 
parameters), and we can use the ones obtained from mean spectral power, with 
necessary spatial and spectral resolution, using different techniques, for example
alternating codes [{\it Lehtinen}, 1986]. 
But the new 'additional' parameters can be determined from single realizations only.

The aim of this paper is to find out the functional dependence of single realization 
spectrum on all the parameters, including well known (temperatures and ion 
composition) and  new ones, which could describe the single realizations spectrum 
properties. For this propose we will use for analysis only signals with high 
signal to noise ratio (more than 10), because in this case the noice 
effects could be neglected and the received signal could be supposed
as only IS signal without presence any noice. 

\section{Initial expressions}
To analyze the individual realizations of the scattered signal, it is necessary
to have a convenient expression relating the spectrum of the scattered signal
to the space-time spectrum of dielectric permittivity irregularities without
averaging over realizations. Such an expression for a monostatic experimental
configuration was obtained and analyzed in [{\it Berngardt and Potekhin}, 2000].
It holds true in the far zone of the receive-transmit antenna and, within constant 
factors (unimportant for a subsequent discussion), is

\begin{equation}
\label{eq:RLU}
u(\omega )=\int H(\omega -\nu ,k-2k_{0}-\frac{\nu }{2c})
\frac{g(-\widehat{\textrm{k}})}{k}\widetilde{\epsilon }
(\nu ,\overrightarrow{k})d\nu d\overrightarrow{k}.
\end{equation}

Here \( \widetilde{\epsilon }(\nu ,\overrightarrow{k}) \) - is the space-time
spectrum of dielectric permittivity irregularities; \( H(\omega ,k)\approx \int H(t,r)e^{-i(\omega t+kr)}drdt=\int o(t)a(t-2r/c)e^{-i(\omega t+kr)}dtdr/r \)
- is the narrow-band weight function; \( a(t),o(t) \) - are, respectively,
the sounder signal envelope and the time window of spectral processing; \( g(\widehat{r}) \)
- is the antenna factor which is the product of the antenna patterns by reception
and transmission; \( \widehat{r}=\overrightarrow{r}/r \) - is a unit vector
in a given direction; \( k_{0} \) - the wave number of the sounding wave; \( c \)
-is the velocity of light. 

Suppose that sounding signal and receiving window of spectral processing are
located in time near the moments \( t=T_{1}-T_{0} \) and \( t=T_{1} \) 
respectively and theirs carriers do not intersects (this is the one of the 
radar equation (\ref{eq:RLU}) obtaining conditions 
[{\it Berngardt and Potekhin}, 2000]). 
In this case the carrier of the weight function \( H(t,r) \) is located near 
the \( t=T_{1};r=T_{0}c/2 \). By going in equation (\ref{eq:RLU}) to the spectrums 
calcualated relative to the weight volume center (to remove the oscillated 
multipliers under integral), we obtain (neglecting to the unessentional multiplier): 

\begin{equation}
\label{eq:RLU_mod}
u(\omega )=
\begin{array}[t]{l}
\int H_{1}(\omega -\nu ,k-2k_{0}-\frac{\nu }{2c})
\frac{g(-\widehat{\textrm{k}})}{k} \\
\widetilde{\epsilon }(\nu ,\overrightarrow{k};T_{1}-T_{0}/2,-\widehat{k}T_{0}c/2)d\nu 
d\overrightarrow{k} \\
\end{array}
.
\end{equation}

where \( H_{1}(\omega ,k)\approx H(\omega ,k)e^{ikT_{0}c/2} \) - low oscillating
part of the \( H(\omega ,k) \), corresponding to its calculation relative to
the center of the weight volume \( H(t,r) \); and \( \widetilde{\epsilon }(\nu ,\overrightarrow{k};T,\overrightarrow{R})=\widetilde{\epsilon }(\nu ,\overrightarrow{k})e^{i(\nu T-\overrightarrow{k}\overrightarrow{R})} \)
- is a time-spatial spectrum of dielectric permittivity irregularities calculated
relative to the point \( t=T,\, \overrightarrow{r}=\overrightarrow{R} \).

In accordance with [{\it Sheffield},1975; {\it Clemmow and Dougherty},1969; 
and {\it Akhieszer et al.},1974], assume that the spectrum of small-scale 
dielectric permittivity irregularities is determined by the Landau solution 
[{\it Landau},1949]. Then the
low-frequency (ion-acoustic) part of the irregularities spectrum in a statistically
homogeneous, unmagnetized, collisionless ionospheric plasma with one sort of
ions is determined by plasma macroscopic parameters (electron and ion temperatures,
ion composition, and drift velocity), and by unknown conditions in the moment
\( T \) related to which this spectrum is calculated - the ions number phase
density in a six-dimensional phase space of velocities and positions of particles.
It is known that the dielectric permittivity irregularities spectrum at large
wave numbers of the sounding wave \( k_{0}>\omega _{N} /c  \) (where \( \omega _{N} \) is plasma frequency)
, is proportional to the electron density irregularities spectrum [{\it Landau and Lifshitz},1982, par.78]:

\[
\widetilde{\epsilon }(\omega ,\overrightarrow{k};T)=-\frac{4\pi q_{e}^{2}}{k_{0}^{2}m_{e}c^{2}}n_{e1}(\omega ,\overrightarrow{k};T),\]

which is given by the expression (for example, [{\it Sheffield}, 1975, sect.6]):

\begin{equation}
\label{eq:Irregularities}
n_{e1}(\omega ,\overrightarrow{k};T)=\frac{G_{e}(\omega ,\overrightarrow{k})}{\epsilon _{||}(\omega ,\overrightarrow{k})}\int \frac{exp(i\overrightarrow{k}\overrightarrow{r})f_{i1}(\overrightarrow{r},\overrightarrow{v};T)}{\omega -\overrightarrow{k}\overrightarrow{v}-i\gamma }d\overrightarrow{r}d\overrightarrow{v},
\end{equation}

where \( \epsilon _{||}(\omega ,\overrightarrow{k})=\left( 1+G_{e}(\omega ,\overrightarrow{k})+G_{i}(\omega ,\overrightarrow{k})\right)  \)
- is longitudinal dielectric permittivity; wave number \( k \) should be small enought 
to wave length be smaller than Debye length (Solpiter approximation). Most part of 
IS radars have the sounding frequencies(50-1000 MHz) within these limitations.

{\par\centering \begin{equation}
\label{eq:G_function}
G_{e,i}(\omega ,\overrightarrow{k})=\frac{4\pi |q_{e,i}q_{e}|n_{e,i0}}{m_{e,i}k^{2}}
\begin{array}{c}
+\infty \\
\int \\
-\infty 
\end{array}\frac{\overrightarrow{k}\frac{\partial f_{0e,i}}{\partial \overrightarrow{v}}}{\omega -\overrightarrow{k}\overrightarrow{v}-i\gamma }d\overrightarrow{v};
\end{equation}
\par}

\( f_{e,i0}(\overrightarrow{v}),\, n_{e,i0} \)- are equilibrium distribution
functions of the electrons and ions velocity and their densities; \( m_{e,i},q_{e,i} \)-
are the mass and charges of electrons and ions, respectively; \begin{equation}
\label{eq:kinetic_approach}
f_{i1}(\overrightarrow{r},\overrightarrow{v};T)=\begin{array}{c}
N\\
\sum \\
j=1
\end{array}\delta (\overrightarrow{r}-\overrightarrow{r_{j}}(T))\delta (\overrightarrow{v}-\overrightarrow{v_{j}}(T))-f_{i0}(\overrightarrow{v})
\end{equation}
 - the ions number phase density in a six-dimensional phase space of velocities
and positions of particles (the ions number phase density, INPD, at \( t=T \)),
with the summation made over all ions. Generally equilibrium distribution functions
\( f_{e,i0}(\overrightarrow{v}) \) are taken to be Maxwellian, with the temperatures
\( T_{e} \) and \( T_{i} \) for electrons and ions, respectively, and in the
absence of a drift they are \[
f_{e,i0}=exp(-(v/V_{T_{e,i}})^{2})/(\pi V^{2}_{T_{e,i}})^{3/2},\]
where \( v_{T_{e,i}}=\left( 2kT_{e,i}/m_{e,i}\right) ^{1/2} \) stands for the
thermal velocities of electrons and ions, respectively. Then the functions \( G_{e,i}(\omega ,\overrightarrow{k}) \)
have the well-known analytical expression, for example [{\it Sheffield},1975]: 

\begin{equation}
\label{eq:G_funct_maxwell}
G_{e,i}(\omega ,\overrightarrow{k})=\left( \frac{1}{k\lambda _{D}}\right) ^{2}\frac{q_{e,i}T_{e}}{q_{e}T_{e,i}}\left( Rw(x_{e,i})-iIw(x_{e,i})\right) 
\end{equation}
where\[
\begin{array}{c}
x_{e,i}=\omega /(kv_{T_{e,i}}); \\
Rw(x)=1-2xe^{-x^{2}}
\begin{array}{c}
x \\
\int \\ 
0 \\
\end{array}
e^{p^{2}}dp \\
Iw(x)=\pi ^{1/2}xe^{-x^{2}} \\
\end{array}
\]
 The physical meaning of the expression (\ref{eq:Irregularities}) is as follows:
the position and velocity of each ion at the moment \( T \) are determined
by the INPD \( f_{i1}(\overrightarrow{r},\overrightarrow{v};T) \), and the
dielectric permittivity irregularities \( \widetilde{\epsilon }(\omega ,\overrightarrow{k};T) \)
are determined by ion-acoustic oscillations of plasma under the action of such
initial conditions. 

\section{Traditional processing of the incoherent scatter signal, and characteristics
of its separate realizations }

Traditionally, the incoherent scattered signal is processed in the following
way. A set of the scattered signal spectra (\ref{eq:RLU}) is used to obtain
its spectral power averaged over realizations. By assuming that an averaging
over the realizations is equivalent to a statistical averaging, and also by
assuming a Maxwellian distribution of the INPD \( f_{i1}(\overrightarrow{r}, \overrightarrow{v};T) \),
one can obtain the following expression for the mean spectral power of the scattered
signal [{\it Suni et al.},1989]: 

\begin{equation}
\label{eq:average_RLU}
<|u(\omega )|^{2}>\approx \frac{const}{R^{2}}\frac{2\pi q_{i}}{kq_{e}}\int F(\omega -\nu )\left| \frac{G_{e}(\nu ,2k_{0})}{\epsilon _{||}(\nu ,2k_{0})}\right| ^{2}f_{i0}(\frac{\nu }{2k_{0}})d\nu 
\end{equation}

where \( F(\omega ) \) is the 'smearing' function determined by the spectrum
of the sounder signal and the spectral processing time window;
and \( <> \) is averaging over realizations. 

The frequency dependence of the scattered signal mean spectral power (\ref{eq:average_RLU})
under usual ionospheric conditions has a typical 'two-hump' form (Figure~\ref{figone}(a)) 
[{\it Evans},1969].
From the scattered signal mean spectral power (\ref{eq:average_RLU}) it is
possible to determine the electron \( T_{e} \) and ion \( T_{i} \) temperatures
and the drift velocity \( \overrightarrow{v}_{0} \) involved in a familiar
way in the functions \( \epsilon _{||}(\omega ,2k_{0}),G_{e}(\omega ,2k_{0}) \)
and \( f_{i0}(v) \) [{\it Sheffield},1975].

In single realizations, however, the scattered signal spectral power differs
essentially from the mean spectral power. Figure~\ref{figone} presents 
the scattered signalspectral power in three consecutive realizations 
(Figure~\ref{figone}(A-C)), the spectral power
averaged over 1000 realizations (Figure~\ref{figone}(D)), and the 
spectral power of the sounder signal envelope (Figure~\ref{figone}(E)). 
From Figure~\ref{figone} it is evident that the non averaged spectral
power of the incoherent scatter signal (Figure~\ref{figone}(A-C)) 
has a typical 'peaked' form, the width of peaks is larger than that of the 
sounder signal spectrum, and the peaks themselves are concentrated in the band 
of the mean signal spectral power. In the case of an averaging over realizations, 
such a peaked structure transforms to a typical smooth two-hump structure 
(Figure~\ref{figone}(D)).

\section{Model of single realizations of incoherent scatter signals}

\subsection{Structure of one realization of the incoherent scatter signal spectrum }

To obtain a model of the incoherent scatter signal we substitute the Landau's
expression for dielectric permittivity irregularities (\ref{eq:Irregularities})
into the expression for the scattered signal spectrum (\ref{eq:RLU_mod}). Using
in (\ref{eq:Irregularities}) the spatial spectrum \( \widetilde{f_{i1}}(\overrightarrow{k},\overrightarrow{v};T) \)
of the ions number phase density \( f_{i1}(\overrightarrow{r},\overrightarrow{v};T) \),
and upon interchanging the order of integration, we obtain the one-realization
model for the scattered signal spectrum: \begin{equation}
\label{eq:signal_model}
u(\omega )=\int K(\omega ,\overrightarrow{k},v_{||})F_{i1}(\overrightarrow{k},v_{||};T_{1}-T_{0}/2)d\overrightarrow{k}dv_{||},
\end{equation}

{\par\centering \begin{equation}
\label{eq:kern_model}
K(\omega ,\overrightarrow{k},v_{||})=\frac{g(-\widehat{k})}{k^{3}}\int \xi (\nu ,\overrightarrow{k})\frac{H(\omega -\nu ,k-2k_{0}-\nu /c)}{(\nu -kv_{||}-i\gamma )}d\nu .
\end{equation}
\par}

Here\begin{equation}
\label{eq:f_integral}
F_{i1}(\overrightarrow{k},v_{||};T)=\int \widetilde{f}_{i1}(\overrightarrow{k},\overrightarrow{v};T)\delta (\overrightarrow{k}\overrightarrow{v}-kv_{||})d\overrightarrow{v}
\end{equation}
is unknown function we want to determine from experiment and has a form similar
to the Radon transform of the function \( \widetilde{f}_{i1}(\overrightarrow{k},\overrightarrow{v};T) \).
A kinetic function (showed at Figure~\ref{figtwo}) 
\begin{equation} 
\xi (\nu ,\overrightarrow{k})=G_{e}(\nu ,\overrightarrow{k})/\epsilon _{||}(\nu ,\overrightarrow{k}) 
\end{equation}
is determined by macroscopic parameters of ionospheric plasma \( T_{e,i} \)
and \( \overrightarrow{v_{0}} \); these parameters can be determined, for example, 
from measurements of the mean spectral power of the received signal 
(\ref{eq:average_RLU}).

Thus the kernel \( K(\omega ,\overrightarrow{k},v_{||}) \) is completely determined
by the sounder signal, the receiving window, and by macroscopic characteristics
of ionospheric plasma. The expression (\ref{eq:signal_model}) clearly
shows the meaning of the kernel \( K(\omega ,\overrightarrow{k},v_{||}) \):
it determines the selective properties of the model, i.e. the possibilities
of determining the unknown function \( F_{i1}(\overrightarrow{k},v_{||};T) \)
from the measured \( u(\omega ) \). Hence it can be termed the weight volume
in the space \( (\overrightarrow{k},v_{||}) \), or ambiguity function. Since
the function \( H_{1}(\omega ,k) \) is a narrow-band one, with its carrier
concentrated near zero, the function of indefiniteness \( K(\omega ,k,v_{||}) \)
has also a limited carrier in \( k \) near \( k=2k_{0} \). The possibilities
of determining the unknown function \( F_{i1}(\overrightarrow{k},v_{||};T) \)
dependence on the wave vector directions \( \widehat{k} \) are determined by
the product of the kinetic function \( \xi (\omega ,\overrightarrow{k}) \)
and the antenna beam \( g(-\widehat{k}) \).

According to the resulting model (\ref{eq:signal_model}), the form of scattered
signal single spectrum is determined both by a determinate component (the weight
volume \( K(\omega ,k,v_{||}) \)), and by a random (i.e. dependent on time
by the unknown way) component. A random component is the function 
\( F_{i1}(\overrightarrow{k},v_{||};T) \)
determined by the spartial harmonics packet of the INPD 
\( \widetilde{f}_{i1}(\overrightarrow{k},\overrightarrow{v};T) \)
with wave numbers \( k \), concentrated near \( 2k_{0} \) and calculated relative
to the moment \( t=T \). The moment \( T \) is determined by the moments of
the sounding signal transmitting and spectral processing receiving window location,
and corresponds to the middle moment between them \( T=T_{1}-T_{0}/2 \). The
weight volume \( K(\omega ,k,v_{||}) \) determines the parameters of this wave
packet, the region of wave vectors and velocities in \( F_{i1}(\overrightarrow{k},v_{||};T) \)
which make the main contribution to the scattered signal at a particular frequency
\( \omega  \).

\subsection{Qualitative properties of the weight volume }

\begin{figure}
\vskip1in 
\resizebox*{0.9\textwidth}{!}{
\includegraphics{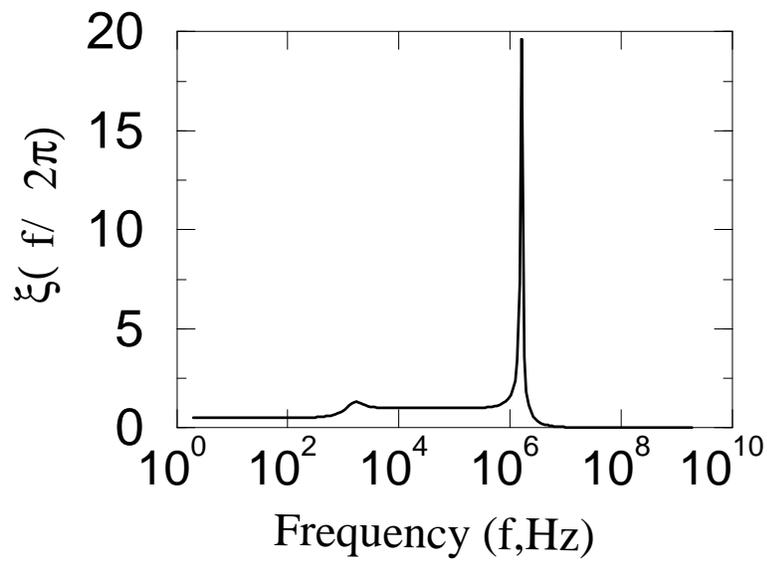} 
}
\caption {Kinetic function \( \xi ( \nu ,\vec{k} ) \) as a function \
 of \( \nu \) at a fixed \( k=2k_{0}=6.28 m^{-1} \) for a typical ionospheric \
 plasma with \( O^{+} \) ions and \( T_{e}=T_{i}=1500K \). 
 } 

\label{figtwo}
\end{figure}

In this experiment the spectra sounder signal \( a(\omega ) \) and the receiving 
window \( o(\omega ) \)
are selected such as they are sufficiently narrow-band ones (1KHz), in comparison 
with the functions \( \epsilon _{||}(\omega ,2k_{0}),G_{e}(\omega ,2k_{0}) \) and
\( f_{i0}(v) \), in order to improve the accuracy of their determination from
experimental results. Therefore, the weight function \( H_{1}(\omega ,k)\sim o(\omega -ck/2)a(ck/2) \)
can also be considered narrow-band from both arguments as compared with the
kinetic function \( \xi (\nu ,\overrightarrow{k}) \) from corresponding arguments.
In this case we can approximate the weight volume 
\( K(\omega ,\overrightarrow{k},v_{||}) \) as:
\begin{equation}
\label{eq:kern_model_modified}
K(\omega ,\overrightarrow{k},v_{||})\approx \frac{\xi (\omega ,2k_{0}\widehat{k})g(-\widehat{k})}{k^{3}}\int \frac{H_{1}(\omega -\nu ,k-2k_{0}-\nu /c)}{(\nu -kv_{||}-i\gamma )}d\nu .
\end{equation}
 The function \( H_{1}(\omega ,k) \) is concentrated near \( \omega =0,\, k=0 \)
[{\it Berngardt and Potekhin},2000]. The width of this function from arguments \( (\omega ,k) \) is
\( \Delta \omega =(\Delta \omega _{a}+\Delta \omega _{o}) \), \( \Delta k=(\Delta \omega _{a}+\Delta \omega _{o})/c \),
where \( \Delta \omega _{a},\Delta \omega _{o} \), is the width of bands of
sounder signal spectra and of the spectral processing window, respectively.
In this experiment on ionospheric sounding by the incoherent scatter
method, they have the order \( \Delta \omega _{a,o}\sim 10^{4}sec^{-1} \).
The function \( \xi (\nu ,\overrightarrow{k}) \) at a fixed \( k=2k_{0}=6.28m^{-1} \)
for a typical ionospheric plasma with \( O^{+} \) ions and \( T_{e}=T_{i}=1500K \)
is presented in Figure~\ref{figtwo}. The figure shows that the function \( \xi (\nu ,\overrightarrow{k}) \)
does varies smoothly on the characteristic size of the weight function 
\( H_{1}(\omega ,k) \) carrier in \( \omega  \) which is in this case has
the order of \( \Delta \Omega \sim 10^{4}sec^{-1} \) (corresponds to a sounding
by the impulse radio signal of a duration of 1 millisecond). 

Assuming that the envelope of the sounder pulse \( a(t) \) and the receiving
window \( o(t) \) have an identical Gaussian-like spectrum: \( a(\omega )=o(\omega )=exp(-(\omega /\Delta \omega )^{2}) \),
we obtain the function \( H_{1}(\omega ,k) \) of the form \begin{equation}
\label{eq:model1}
H_{1}(\omega ,k)=Be^{-((\omega -kc/2)/\Delta \omega )^{2}}e^{-(kc/(2\Delta \omega ))^{2}}.
\end{equation}
 Upon substituting (\ref{eq:model1}) into (\ref{eq:kern_model_modified}) and
rather unwieldy calculations, similar to [{\it Landau}, 1946], we obtain the 
following expression for the scattered signal spectrum (\ref{eq:signal_model}): 

\begin{equation}
\label{eq:simpl_model}
u(\omega )=\int K(\omega ,\overrightarrow{k},v_{||})\widetilde{F}_{i1}(\overrightarrow{k},v_{||};T_{1}-T_{0}/2)dv_{||}d\overrightarrow{k},
\end{equation}
where \begin{equation}
\label{eq:modified_kernel}
K(\omega ,\overrightarrow{k},v_{||})=V_{1}(\omega -kv_{||})V_{2}(\omega -(k-2k_{0})c)V_{3}(\omega ,\widehat{k})
\end{equation}
\begin{equation}
\label{eq:K2_form}
\begin{array}{c}
V_{1}(\omega )=i\pi e^{-\Phi ^{2}(\omega )}-Rw(\Phi (\omega )); \\
V_{2}(\omega )=e^{-\Phi ^{2}(\omega )}; \\
V_{3}(\omega ,\widehat{k})=\frac{\xi (\omega ,2k_{0}\widehat{k})g(-\widehat{k})}{8k_{0}^{3}} \\
\end{array}
\end{equation}
\begin{equation}
\label{eq:Fases}
\Phi (\omega )=\frac{\omega }{\sqrt{2}\Delta \omega }
\end{equation}
The selective properties of \( K(\omega ,\overrightarrow{k},v_{||}) \) in the
longitudinal component of the velocity \( v_{||} \) are determined by the first
cofactor \( V_{1} \) in (\ref{eq:modified_kernel}). A maximum 
\( K(\omega ,\overrightarrow{k},v_{||}) \)
in \( v_{||} \) at a fixed \( \omega  \) is determined by the condition 
\( V_{1}(\omega -kv_{||,0})=max \)
which, as a consequence of the properties of the exponential and the \( Rw \)
functions, corresponds to the frequency Doppler shift condition in the case
of the scattering from a single particle: 

\begin{equation}
\label{eq:drift_w}
v_{||,0}=\frac{\omega }{k}.
\end{equation}
 The width \( \Delta v_{||} \) of a maximum \( K(\omega ,\overrightarrow{k},v_{||}) \)
in \( v_{||} \) (that determines the region of velocities making the main contribution
to the scattered signal at fixed \( \omega  \) and \( \overrightarrow{k} \))
can be estimated from the condition \( \Phi (\omega -kv_{||,0}\pm \Delta v_{||})=1 \)
to be \begin{equation}
\label{eq:accurcy_v}
\Delta v_{||}=\sqrt{2}\Delta \omega /k.
\end{equation}
 The selective properties of \( K(\omega ,\overrightarrow{k},v_{||}) \) in
wave numbers \( k \) are determined by the second cofactor \( V_{2} \) in
(\ref{eq:modified_kernel}). A maximum \( K(\omega ,\overrightarrow{k},v_{||}) \)
in \( k \), at a fixed \( \omega  \), is determined by the condition \( V_{2}(\omega -(k-2k_{0})c)=max \)
which corresponds to the condition (analogical to the Volf-Bragg condition for
scattering from nonstationary spatial harmonic):\begin{equation}
\label{eq:k_w}
k=2k_{0}+\frac{\omega }{c}.
\end{equation}
 The width \( \Delta k \) of a maximum \( K(\omega ,\overrightarrow{k},v_{||}) \)
in \( k \) (that determines the region of wave numbers making the main contribution
to the scattered signal at a fixed \( \omega  \)) can be estimated from the
condition \( \Phi (\omega -(k-2k_{0})c\pm \Delta kc)=1 \) to be \begin{equation}
\label{eq:accuracy_k}
\Delta k=\sqrt{2}\Delta \omega /c.
\end{equation}
 The function \( V_{3} \) determines the selective properties of \( K(\omega ,\overrightarrow{k},v_{||}) \)
in the direction of the wave vectors \( \widehat{k} \) and the maximum possible
width of the received signal spectrum determined by the kinetic function 
\( \xi (\omega ,2k_{0}\widehat{k}) \) and antenna factor \( g(-\widehat{k}) \). 

\section{Discussion }

From (\ref{eq:simpl_model}) it follows that the fine structure of the scattered
signal spectrum \( u(\omega ) \) is determined only by the properties of the
function \( \widetilde{F}_{i1}(\overrightarrow{k},v_{||};T) \) (\ref{eq:f_integral})
and can be related to the INPD \( \widetilde{f}_{i1}(\overrightarrow{k},\overrightarrow{v};T) \)
only within the framework of additional assumptions about the structure of 
\( \widetilde{f}_{i1}(\overrightarrow{k},\overrightarrow{v};T) \).
Formally, to determine the function \( \widetilde{f}_{i1}(\overrightarrow{k},\overrightarrow{v};T) \)
require measuring the scattered signal for different wave numbers \( k_{0} \)
of the sounder signal simultaneously. 

To carry out a qualitative comparison with experimental spectra of incoherently-scattered 
signals we use the spectral processing window that repeats the sounder signal
shape. Their spectra are approximated by Gaussian-like spectra with a width
equal to their actual spectrum width. According to (\ref{eq:simpl_model}),
the spectrum of the received signal is defined by the unknown function \( \widetilde{F}_{i1}(\overrightarrow{k},v_{||};T) \)(\ref{eq:f_integral})
(convoluted with a kernel \( K(\omega ,\overrightarrow{k},v_{||}) \)).
For comparison with experimental data, we give following simple model of the
function \( \widetilde{F}_{i1}(\overrightarrow{k},v_{||};T) \). 

\textbf{Simple model} Assume that \( \widetilde{f}_{i1}(\overrightarrow{k},\overrightarrow{v};T) \)
involves only one peak in the range of longitudinal wave vectors and velocities
of our interest:
\[
\widetilde{f}_{i1}(\overrightarrow{k},\overrightarrow{v};T)=\delta (\overrightarrow{k}-\overrightarrow{k_{1}})\delta (\overrightarrow{v}-\overrightarrow{v_{1}}).\]
\[
\widetilde{F}_{i1}(\overrightarrow{k},v_{||};T)
=\delta (\overrightarrow{k}-\overrightarrow{k_{1}})
\delta (\overrightarrow{k}\overrightarrow{v_{1}}-kv_{||}).\]
This model corresponds to the fact that the medium involves an isolated spatial
harmonic \( \widetilde{f}_{i1}(\overrightarrow{k},\overrightarrow{v};T) \)
at the wave vector \( \overrightarrow{k_{1}} \) and with longitudinal velocity
\( v_{1} \), the amplitude of which is significantly larger than the amplitudes
of spatial harmonics close to it. The spectrum of the received signal will then
involve also only one peak:\begin{equation}
\label{eq:rez_model2}
u(\omega )\sim V_{1}(\omega -k_{1}v_{1})V_{2}(\omega -(k_{1}-2k_{0})c)V_{3}(\omega ,\widehat{k_{1}}),
\end{equation}
 and the form and width of this peak will be defined by the product \( V_{1}(\omega -k_{1}v_{1})V_{2}(\omega -(k_{1}-2k_{0})c) \).
From the position of the peak in the spectrum \( u(\omega ) \), one can determine
the unknown wave number \( k_{1} \) of the spatial harmonic \( \widetilde{f}_{i1}(\overrightarrow{k},\overrightarrow{v};T) \)
that make the main contribution to the scattered signal (\ref{eq:k_w}) and
the longitudinal velocity \( v_{1} \) (\ref{eq:drift_w}) corresponding to
the moment \( T \) . 

This model gives the signal spectrum with only one peak. According to the 
model, the observed presence of several peaks in the real spectrum corresponds 
to existance not only one peculiarity described by the model but a numer of them.

Let us show that after averaging this model gives as the well known expression 
for the IS average spectral power. According traditional approaches lets suppose 
that \( <f_{i1}(\vec{k},\vec{v},T)>=f_{i0}(\vec{v}) (when |k|>0) \).
Summarizing over the different realizations gives the produces the function:\begin{equation}
\label{eq:mod1_rezult}
<|u(\omega )|^{2}>\sim  \begin{array}[t]{l}
\int \left| V_{3}(\omega ,\widehat{k}) \right| ^{2} d\widehat{k} 
\sum_{j=1}^{N} f_{i0}(v_{||,j})\\
\left| V_{1}(\omega -2k_{0}v_{||,j})\right| ^{2} \int \left| V_{2}(\omega -(k-2k_{0})c)\right| ^{2} dk\\
 =const \left| \int V_{3}(\omega ,\widehat{k})d\widehat{k}\right| ^{2}\int f_{i0}(v)\left| V_{1}(\omega -2k_{0}v)\right| ^{2}dv\\
\sim \left| \frac{G_{e}(\omega ,2k_{0})g(-\widehat{k_{0}})}{8k_{0}^{3}\epsilon _{||}(\omega ,2k_{0})}\right| ^{2}\int f_{i0}(v)\left| V_{1}(\omega -2k_{0}v)\right| ^{2}dv
\end{array}
\end{equation}

From (\ref{eq:mod1_rezult}) becomes clean, that the average received signal spectral 
power \( <|u(\omega )|^{2}> \) is determined as a product of the kinetic function 
\( |\xi (\omega ,2k_{0}\widehat{k})|^{2} \) and maxwellian ions distribution 
\( f_{i0}(v) \) convolved with a narrow-band function 
\( \left| V_{1}(\omega -2k_{0}v)\right| ^{2} \) (which is defined by
the sounding signal and spectral processing receiving window spectra). Qualitatively
this solution (\ref{eq:mod1_rezult}) is close to the average spectral power
obtainded by the tradditional way (\ref{eq:average_RLU}). 

\begin{figure*}
\vskip1in 
\resizebox*{0.9\textwidth}{!}{
\includegraphics{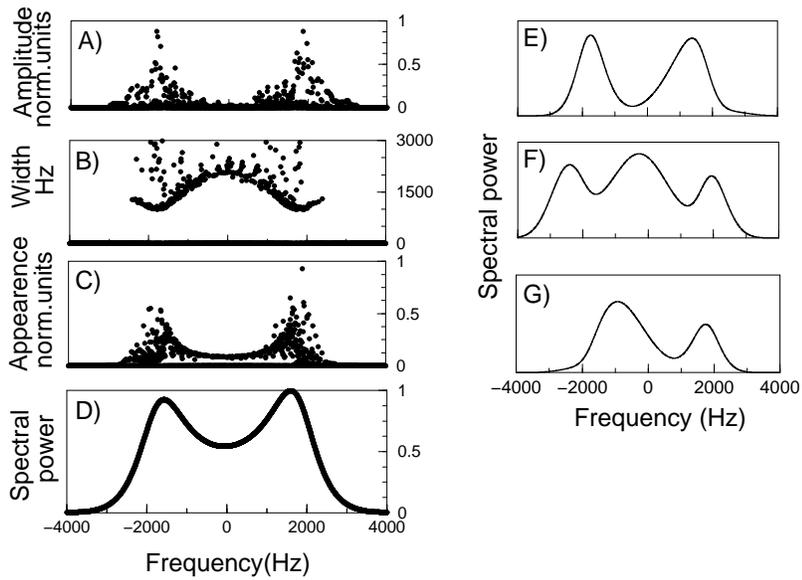}
}
\caption{ Numerical simulation results for \( T_{e}=T_{i}=1500K \) 
using equations (\ref{eq:signal_model}), (\ref{eq:modified_kernel}) 
and simple model (\ref{eq:comp_model}). 
Mean peak amplitude (A),mean peak width(B), mean peak appearence(C),  
average spectral power (D) and three spectral realizations (E-G).
} 
\label{fig4}
\end{figure*}

A numerical simulation have been made for the model containing a number of peaks:
\begin{equation}
\label{eq:comp_model}
\widetilde{F}_{i1}(\overrightarrow{k},v_{||};T)
=\delta(sin(kA+\phi(T))*sin(B(v_{||})+\phi(T))-1),
\end{equation}
where 
\( A \) - some constant;
\( B(v) \sim \int_{0}^{v} exp(-x^{2}/v_{Ti}^{2})dx \); 
\( \phi(T) \) - a function with uniform distribution of values (we have used 
random one, having fixed value for fixed \( T \) ).
This expression for \( F_{i1} \) has a normal distribution over 
\( v_{||} \), uniform distribution over \( k \), and statistical independence 
of the values for different \( T \) (or when delay between them exceeds the 
interval in which \( \phi(T) \) changes slowly).
Spectra obtained by substituting this simple model (\ref{eq:comp_model}) into 
obtained equation (\ref{eq:signal_model}), gives us the following single spectrums 
and their statistical properties, are shown at  Figure~\ref{fig4}. 
As one can see, single realizations of the spectral power(E-G) have the same 
structure, as an experimental one (Figure~\ref{figone}), 
the close peak width (B) (Figure~\ref{figstat},B), and same 
relation between mean peak appearence (C) and average spectral power (D) 
(Figure~\ref{figstat},C,D). This allows us to suppose that the 
model of the scattered signal single spectrum (\ref{eq:signal_model}) could 
be used to describe signal properties, and simplified model (\ref{eq:comp_model}) 
could quilitatively describe the ion number phase density behavior. 

\section{Conclusion }

In this paper we have suggested an interpretation of separate realizations of
incoherently scattered signal spectra. It is based on the radar equation 
[{\it Berngardt and Potekhin},2000]
and kinetic theory of ion-acoustic oscillations of a statistically homogeneous
unmagnetized, collisionless ionospheric plasma with one sort of ions 
[{\it Landau},1946].

In accordance with the proposed model (\ref{eq:signal_model}), the main contribution
to the scattering is made by plasma waves caused by spatial harmonics of ions
number phase density \( \widetilde{f}_{i1}(\overrightarrow{k},\overrightarrow{v};T) \),
with wave numbers on the order of the double wave number of the sounder signal
\( k\approx 2k_{0} \), for \( T=T_{1}-T_{0}/2 \). 

It has been shown that the form of the received signal spectrum \( u(\omega ) \)
is related to the INPD \( \widetilde{f}_{i1}(\overrightarrow{k},\overrightarrow{v};T) \).
At each frequency \( \omega  \) the value of \( u(\omega ) \) is determined
by Radon's-like integral \( \widetilde{F}_{i1}(\overrightarrow{k},v_{||};T) \)
on \( \widetilde{f}_{i1}(\overrightarrow{k},\overrightarrow{v};T) \) between
the limits of the velocity components across the wave vector \( \overrightarrow{k} \)
(\ref{eq:f_integral}). The region of wave vectors and longitudinal velocities
making contribution to the received signal \( \overrightarrow{k},v_{||} \)
is determined by the weight volume \( K(\omega ,\overrightarrow{k},v_{||}) \)
(\ref{eq:kern_model}). 

So, by changing the transmitting pulse start moment and the moment of its receiving
one could measure the value of \( \widetilde{F}_{i1}(\overrightarrow{k},v_{||};T) \),
as function of the time \( T \). This allows the INPD \( \widetilde{f}_{i1}(\overrightarrow{k},\overrightarrow{v};T) \)
diagnostics for different moments including delays smaller than theirs lifetime
and without statistical averaging of receiving signal. Actually, in the case
of irregularities lifetime much longer than sounding pulses repeating interval
\( T_{i+1}-T_{i} \)., one could measure the \( \widetilde{F}_{i1}(\overrightarrow{k},v_{||};T_{i}-T_{0}/2) \)
behavior as function of the time \(t=T_{i}-T_{0}/2 \).

Based on the proposed model in (\ref{eq:signal_model}) and a Gaussian approximation
of the spectra of the sounder signal envelope and the receiving window, a qualitative
comparison of the model with experimental data from the Irkutsk incoherent scatter
radar was carried out. The comparison showed a quilitative agreement for simplified
model (\ref{eq:comp_model}), based on additional assumptions about the properties
of the function \( \widetilde{F}_{i1}(\overrightarrow{k},v_{||};T) \). 

\section*{Acknowledgements}
I am grateful to B.G.Shpynev for making the data from the Irkutsk IS radar available
and to A.P.Potekhin for fruitful descussions. 
The work has been done under partial support of RFBR grants \#00-05-72026 
and \#00-15-98509.

\end{document}